# Microwave Properties of Ba-Substituted Pb(Zr$_{0.52}$Ti$_{0.48}$)O$_3$ after Chemical-Mechanical Polishing

Federica Luciano,[1,2] Lieve Teugels,[2] Sean McMitchell,[2] Giacomo Talmelli,[2] Anaïs Guerenneur,[2,3] Renzo Stheins,[4] Rudy Caluwaerts,[2] Thierry Conard,[2] Inge Vaesen,[2] Stefanie Sergeant,[2] Pol Van Dorpe,[2,3] Stefan De Gendt,[1,2] Matthijn Dekkers,[4] Johan Swerts,[2] Florin Ciubotaru,[2] and Christoph Adelmann[2,*]

[1]KU Leuven, Department of Chemistry, 3001 Leuven, Belgium

[2]Imec, 3001 Leuven, Belgium

[3] KU Leuven, Department of Physics and Astronomy, 3001 Leuven, Belgium

[4]Solmates B.V., 7521 PE Enschede, The Netherlands

**Abstract**

We have studied the effect of chemical-mechanical polishing (CMP) on the ferroelectric, piezoelectric, and microwave dielectric properties of Ba-substituted PZT (BPZT), deposited by pulsed laser deposition. CMP allowed for the reduction of the root mean square surface roughness of 600 nm thick BPZT films from 12.1±0.1 nm to 0.79±0.15 nm. Ammonium peroxide (SC-1) cleaning was effective to remove Si CMP residuals. Measurements of the ferroelectric hysteresis after CMP indicated that the ferroelectric properties of BPZT were only weakly affected by CMP, while the piezoelectric d$_{33}$ coefficient and the microwave permittivity were reduced slightly by 10%. This can be attributed to the formation of a thin dead layer at the BPZT surface. Moreover, the intrinsic dielectric permittivity at microwave frequencies between 1 and 25 GHz was not influenced by CMP, whereas the dead layer series capacitance decreased by 10%. The results indicate that the CMP process can be used to smoothen the BPZT surface without affecting the film properties strongly.

---

[*] Author to whom correspondence should be addressed. Email: christoph.adelmann@imec.be

**Introduction**

In recent years, piezoelectric materials have received growing interest for actuation and sensing in microelectromechanical systems (MEMS).[1,2] Lead zirconate titanate Pb(Zr$_{1-x}$Ti$_x$)O$_3$ (PZT) is among the most studied materials and has been extensively researched due to its strong piezoelectric response, especially near the morphologic phase boundary at $x = 0.48$.[3,4] This has led to an increasing number of applications of PZT in MEMS.[5,6] Specifically, PZT has been considered *e.g.* for ultrasound transducers[7–9] or high-frequency bulk acoustic resonators.[10,11] Furthermore, PZT has been proposed as piezoelectric material in multiferroic composites[12–14] for microelectronic applications.[15–18] However, a key limitation of PZT thin films in such applications can be the reduced dielectric (and therefore piezoelectric) response at high frequencies, especially in the microwave regime, due to dielectric relaxation and large losses.[19,20] This can limit the operating frequency range of PZT-based MEMS and multiferroic devices, particularly in the GHz range.

Recently, it has been reported that Ba-substituted[21] PZT (BPZT) thin films deposited by pulsed laser deposition (PLD) do not show dielectric relaxation up to frequencies of at least 30 GHz, which renders such films promising for RF MEMS and microelectronic applications.[22] Textured polycrystalline BPZT films typically show a columnar growth that improves their out of plane piezoelectric properties.[22,23] However, a drawback of such columnar growth is the large surface roughness, which can be a major obstacle for the integration of PLD BPZT films in MEMS and multiferroic compounds, limiting both performance and reliability.

This issue can be mitigated by chemical-mechanical polishing (CMP) to planarize and smoothen the BPZT layer. Several publications have reported CMP processes and post-CMP cleaning procedures for PZT thin films fabricated by sol-gel techniques[24–26] or MOCVD.[27,28] However, no report has been published on the CMP of BPZT deposited by PLD. Moreover, the effect of the CMP process on the ferroelectric[25,26] and piezoelectric[29] properties of PZT has not been studied comprehensively and the impact on the microwave dielectric properties is still unknown. In this paper, we report on the CMP of BPZT using a silica slurry in combination with an SC-1 post-CMP clean to remove slurry residues. We assess the surface morphology and surface composition and contamination after CMP and cleaning. Finally, we determine the ferroelectric, piezoelectric, and microwave dielectric properties and compare them to as-deposited BPZT. The results indicate that the CMP process strongly reduces the surface roughness while inducing only a slightly enlarged dielectric dead layer at the top surface. Hence, the results indicate that CMP can be used to obtain flat and smooth BPZT films with excellent ferroelectric, piezoelectric, and particularly microwave dielectric properties.

**Experimental details**

*Deposition and characterization techniques.*

All BPZT [$(Ba_{0.1}Pb_{0.9})(Zr_{0.52}Ti_{0.48})O_3$] layers were deposited by pulsed laser deposition (PLD) in a Solmates SIP-700 system on Si (100). Prior to BPZT deposition, a 400 nm thick thermal $SiO_2$ layer was grown, followed by the deposition of a 20/70 nm thick Ti/Pt bottom electrode (BE) by e-beam evaporation. The polycrystalline Pt layer exhibited a strong (111) fiber texture. Subsequently, a 10 nm thick $LaNiO_3$ (LNO) buffer layer was deposited by PLD,[22] which has been shown to improve the dielectric[30] and piezoelectric[31] properties of PZT. Finally, the BPZT layers were deposited by PLD at 550°C from a single compound target in an $O_2$ atmosphere. Fig. 1a shows a schematic of the sample stack.

Film thicknesses *d* were measured using cross-sectional scanning electron microscopy (X-SEM) using a Hitachi SU8000 field-emission microscope. The phase and texture of the BPZT layers were determined by 2θ-ω x-ray diffraction (XRD) in a Malvern Panalytical X'Pert diffractometer using Cu Kα radiation. The surface roughness of the layers was assessed by atomic-force microscopy (AFM) images using a Bruker Dimension Icon microscope in tapping mode. The surface composition was quantified by x-ray photoelectron spectroscopy (XPS) with a Physical Electronics QUANTES instrument in angle-integrated mode using Al Kα radiation.

*CMP and cleaning processes.*

The BPZT samples were polished in a Mecapol E460 system using an Optivision 4548 soft pad and a silica ($SiO_2$) abrasive slurry (ILD3013, pH 10–11). Two conditions were used for the rotation speed of the top (head) and bottom (table) plates as well as for the applied down force. The parameters combinations for the different conditions are listed in Tab. 1.

Post-CMP cleaning was performed by immersion in an SC-1 solution ($NH_4OH:H_2O_2:H_2O$ = 1:1:5) at 75ºC for 10 min followed by a 5 min soak in isopropanol in an ultrasonic bath. The different polishing conditions were evaluated in terms of final surface roughness (by AFM) and CMP removal rate (by X-SEM), the latter by measuring the film thickness before and after CMP and cleaning. Thickness measurements after cleaning only indicated a negligible (<< 1 nm/min) etch rate of BPZT by the SC-1 solution itself.

*Ferroelectric and piezoelectric measurements*

To characterize the ferroelectric and piezoelectric properties of BPZT thin films, square capacitors (1 mm diameter) with 100 nm thick Pt top electrodes were fabricated using photolithography and lift-off processes. The same devices were also used to measure the BPZT piezoelectric coefficient $d_{33}$. The polarization *P* of the BPZT thin films was measured *vs.* external voltages *V* (electric field $E = V/d$) using an aixACCT TF Analyzer 3000 at a frequency of 1 kHz before as well as after CMP and cleaning. Similarly, an aixACCT DBLI system was used to determine the piezoelectric coefficient $d_{33}$ (also at 1 kHz) for the BPZT thin films as deposited as well as after CMP and cleaning.

*Microwave dielectric permittivity measurements*

The microwave dielectric properties of the BPZT films were assessed using concentric capacitor (CC) test structures[22,32] with Ti/Au (10 nm/1.5 µm) top electrodes, patterned by photolithography and lift-off processes. A schematic of the CC test structures is shown in Fig. 1b. The outer pad radius was set to 90 µm, while the inner capacitor radius was varied between 10 µm and 15 µm. Such small inner capacitor diameters were chosen to increase the measurement bandwidth and sensitivity.[11]

All electrical measurements were carried out on a microwave probe station in a one-port measurement setup using a ground-signal-ground (GSG) probe (see Fig. 1b). Microwave S-parameters ($S_{11}$) were measured using an AGT E8363B vector network analyzer (VNA), which was calibrated by a one-port short-open-load approach prior to the measurements. For the large inner capacitances studied here, the dielectric constant of the dielectric film with thickness $d$ can then be extracted from $S_{11}$ measurements of two CC structures with different inner capacitor radii $a_1$ and $a_2$ via[32]

$$\varepsilon_r = \frac{d\left(\frac{1}{a_2^2} - \frac{1}{a_1^2}\right)(X_1 - X_2)}{\omega\pi\varepsilon_0\left\{(X_1 - X_2)^2 + \left[R_1 - R_2 - \frac{R_B}{2\pi}\ln\left(\frac{a_2}{a_1}\right)\right]^2\right\}} \quad [1]$$

Here $Z_i = R_i + jX_i = Z_0 \cdot \frac{1+S_{11,i}}{1-S_{11,i}}$ is the impedance of the device $i$ = 1,2 under test ($Z_0$ = 50 Ω), $\omega = 2\pi f$ is the angular frequency of the measurement, and $\varepsilon_0$ is the vacuum permittivity. $R_b$ is the sheet resistance of the Pt bottom electrode, which was measured by four-point electrical probing to be 2.1 Ω/□.

**Results and discussion**

This section is organized as follows. First, we describe the thin film properties, the CMP and cleaning process results, as well as the BPZT surface properties after CMP and cleaning. Next, the effect of CMP and cleaning on ferroelectric and piezoelectric properties at low frequencies is reported. Finally, the BPZT microwave dielectric properties after CMP and cleaning are discussed.

*Film properties, CMP and cleaning processes, surface roughness and composition*

The 2θ-ω XRD pattern of a 600 nm thick film in Fig. 2a revealed strong (110) fiber texture of the perovskite BPZT layer. No signs of a pyrochlore secondary phase could be detected.

XSEM (Fig. 2b) revealed a columnar microstructure. This columnar growth mechanism and microstructure were principally responsible for the high surface roughness of the BPZT layers, as measured by AFM (Fig. 2c). The extracted root mean square (RMS) roughness of a 600 nm thick BPZT layer was as high as 12.1 nm.

Figure 3 shows AFM images of BPZT surfaces obtained after CMP and cleaning at different conditions, as described in the previous section (see also Tab. 1). In all cases, the RMS roughness was markedly reduced by CMP to values around 1 nm (Tab. 1) for films with post-CMP thicknesses around 450 nm. Lowest RMS values were attained for condition 1 (Fig. 3a, RMS = 0.8 nm) and condition 3 (Fig. 3c, RMS = 0.9 nm). This indicates that low downforce is beneficial for low final surface roughness whereas the rotation speed had less impact. In all cases, the final roughness was reached after 2 minutes of polishing; longer polishing times (Fig. 3b) did not reduce the surface roughness further.

X-SEM measurements of the BPZT thickness before and after CMP were used to calculate the removal rate (RR) for the different polishing conditions. The results are summarized in Tab. 1 and indicate lower RRs with respect to the literature for the same slurry.[24] The data indicate no clear correlation between RR and RMS roughness. However, condition 1 had the lowest RR value, together with the lowest RMS surface roughness. This indicates that condition 1 was optimum in the studied parameter range as it generates the smoothest surfaces for the lowest removed thicknesses.

The composition of the BPZT surface after CMP and cleaning was studied by XPS, comparing spectra for as deposited BPZT with those obtained after CMP and subsequent cleaning in an SC-1 solution. Both the surface contamination and stoichiometry were analyzed. Figure 4 depicts Si 2p photoelectron spectra for the different process conditions described above. While spectra of as deposited BPZT (Fig. 4a) did not show any peaks related to Si, the CMP process induced Si contamination from the silica slurry, as demonstrated by the observation of two Si-related peaks in Fig. 4b. The peak at 103 eV can be attributed to $SiO_2$, whereas the peak at 101 eV stems from a $SiO_x$ suboxide.[33] A more quantitative analysis found that the Si surface concentration was as high as 8 at.% after CMP. Note that the weak peak around a binding energy (BE) of 106 eV does not originate directly from the Si 2p core level but can be rather interpreted as an energy loss peak. These observations indicate considerable Si contamination on the BPZT surface after CMP. However, the Si 2p intensity decreased below the detection limit of around 0.1 at.% after the SC-1 cleaning step, as shown in Fig. 4c. This demonstrates that the SC-1 cleaning step is very effective to reduce the Si surface contamination. No other surface contaminations above XPS detection limits were detected, neither after CMP nor after cleaning.

Table 2 shows the atomic concentration of BPZT components close to the surface, as deduced from Ba $3d_{5/2}$, Pb 4f, Zr 3d, and Ti 2p XPS spectra for the different CMP conditions. The results show that the surface cation composition of BPZT remains unchanged by CMP and post-CMP cleaning within experimental accuracy. Furthermore, the (normalized) XPS spectra in Fig. 5 reveal that for all process conditions, the cations are present in oxidized form. This implies that there is no preferred removal of any component of BPZT thin films by the polishing procedure, as well as no reduction of the oxide. It should however be noted that the Ba $3d_5$ peak in Fig. 5a shows an additional component at higher BE after cleaning. This component can be attributed to Ba-OH that is formed during the cleaning in aqueous SC-1 due to the strong hygroscopicity of BaO.[34]

*Ferroelectric and piezoelectric behavior of BPZT after CMP and cleaning*

To understand how the optimized CMP and cleaning processes affect the ferroelectric and piezoelectric properties of BPZT thin films, we have performed low frequency ferroelectric polarization measurements following the procedures described above. To separate surface and possible bulk effects, two BPZT thicknesses of 200 nm and 450 nm were studied. To reach targeted final thicknesses, given the CMP removal rate, the initial BPZT film thickness was adjusted accordingly. As deposited BZPT films were used as references with thicknesses equal to the final thickness after CMP and cleaning.

Figures 6a and 6b show the hysteretic behavior of the ferroelectric polarization *P vs.* the applied electric field *E* for 200 nm and 450 nm thick BPZT films, both as deposited as well as post CMP and cleaning. For the thinner film of 200 nm final thickness, the saturation polarization was reduced by about 10-15% after CMP and cleaning. By contrast, the saturation of the 450 nm thick film polarization remained similar after CMP and cleaning within experimental precision. The increased asymmetry of the *P(E)* loops after CMP and cleaning indicates the degradation of the interface with the top electrode. These findings thus suggest that the CMP process leads to a surface layer with reduced permittivity, which then has more impact for thinner BPZT films. Nonetheless, the impact of CMP and cleaning on the ferroelectric behavior appears limited and films remain ferroelectric in bulk. Results for saturation polarization and coercive field both for as deposited as well as CMPed and cleaned BPZT films are summarized in Tab. 3.

Measurements of the piezoelectric $d_{33}$ coefficient paint a similar picture. A reduction of $d_{33}$ by about 25% was found for the 200 nm thick film after CMP and cleaning. By contrast, a smaller reduction of $d_{33}$ by about 15% was found for thicker 450 nm film. These findings are in qualitative agreement with the polarization measurements above and further suggest the formation of a surface dead layer by the CMP process. This dead layer behaves as a series capacitor, reducing the electric field in the piezoelectric material for a given voltage. As consistently confirmed by the experimental results, this dead layer has a growing influence as the thickness of the BPZT is reduced.

To understand whether the surface dead layer was introduced by the CMP or the cleaning process, we have also studied the ferroelectric response of BPZT after cleaning only. The measurements (data not shown) exhibited no effect of the SC-1 cleaning step on ferroelectricity. Hence, the presence of Ba-OH that was revealed by XPS had no impact. As a consequence, chemical modifications of the surface cannot explain the observations. The surface dead layer thus likely stems from amorphization of the surface region by CMP. This surface dead layer may be visible in X-SEM image in Fig. 5e, which shows a thin region with difference contrast close to the top surface.

*Microwave properties of BPZT after CMP and cleaning*

Microwave permittivities were determined using CC test structures for BPZT films both with as well as without CMP and cleaning prior to test structure processing. The permittivity was measured in a frequency range between 1 and 25 GHz for BPZT film thicknesses between 200 nm and 1 μm using the methodology described above. The dielectric permittivity $\varepsilon_r$ was then determined using Eq. (1). The results of $\varepsilon_r$ are shown as a function of frequency for as-deposited BPZT as well as for BPZT after CMP and cleaning in Figs. 7a and 7b, respectively.

The data indicate that the CMP procedure had only a minor impact on the BPZT dielectric characteristics since values and trends with frequency of the dielectric permittivity were essentially the same for as-deposited films and films after polishing and cleaning. For all analyzed samples, the extracted dielectric permittivity at 1 GHz increased from about 800 for the thinnest films to 1300 (for as-deposited BPZT) or 1220 (for BPZT after CMP and cleaning). Furthermore, for both BPZT with and without CMP, $\varepsilon_r$ showed only weak relaxation by about 20-30% in the frequency range up to 20 GHz. This behavior is consistent with a prior report on the dielectric characteristics of BPZT deposited by PLD at microwave frequencies.[11,22] The apparent thickness dependence of $\varepsilon_r$ can be attributed to the presence of a dead layer with a low permittivity at the electrode-ferroelectric interface that forms a series capacitor with the "bulk" BZPT capacitance $C_{BPZT}$.

To separate dead layer and "bulk BPZT" contributions to the capacitance more quantitatively, both for BPZT prior and post CMP, the intrinsic permittivity $\varepsilon_{BPZT}$ can be extracted from the measurements by assuming that the dead layer capacitance, $C_{DL}$, is independent of the BPZT thickness and determined by the dead layer thickness $t_{DL}$ and dielectric permittivity $\varepsilon_{DL}$. Then

$$\frac{1}{C} = \frac{1}{C_{BPZT}} + \frac{1}{C_{DL}} \rightarrow \frac{t}{\varepsilon_r} = \frac{t - t_{DL}}{\varepsilon_{BPZT}} + \frac{t_{DL}}{\varepsilon_{DL}} \qquad [2]$$

with the total stack with final thickness $t = t_{DL} + t_{BPZT}$ and the effective permittivity $\varepsilon_r$, obtained from CC measurements.

The expected linear dependance of $t/\varepsilon_r$ on the thickness $t$ for BPZT at 1 GHz was confirmed by the data in Fig. 7c in the thickness range above 450 nm. From the slope and the intercept of this linear fit, the intrinsic permittivity $\varepsilon_{BPZT}$ and the ratio $t_{DL}/\varepsilon_{DL}$ were extracted, respectively (see Tab. 4). The intrinsic permittivity of both as-deposited and post CMP BPZT was identical within experimental precision and fell within the range between 900 and 2000 reported in previous studies.[35,36] By contrast, the values of $t_{DL}/\varepsilon_{DL}$ (*i.e.*, $1/C_{DL}$) of post CMP BPZT were about 10% larger than for as-deposited films. However, they were still lower than what previously reported,[22] which confirms the low damage of the BPZT surface post CMP. We note that thinner films of 200 nm thickness showed an even lower value of $t_{DL}/\varepsilon_{DL}$ for the extracted value of $\varepsilon_{BPZT}$, which is consistent with the slightly shorter CMP time and an expected resulting smaller absolute surface dead layer capacitance.

These findings are thus consistent with the formation of an additional dead layer on the top surface of the sample after CMP (and cleaning). However, its impact on the dielectric permittivity at microwave frequencies is limited to similar magnitudes than observed for the ferroelectric and piezoelectric response degradation. This indicates that the CMP process is not only preserving well ferroelectric and piezoelectric properties at low frequencies, but especially the microwave dielectric properties up to frequencies of at least 30 GHz.

**Conclusions**

In summary, we have assessed the effect of CMP and post-CMP cleaning on the ferroelectric, piezoelectric, and (microwave) dielectric BPZT thin films. CMP was performed using a silica-based slurry with a soft pad and post-CMP cleaning employed an SC-1 solution. The study of the effect of the CMP parameters on removal rate and surface roughness led to an optimum CMP process that reduced the RMS surface roughness from about 12 nm to 0.8 nm after 2 min of polishing at a removal rate of 1.6 nm/s. XPS revealed no (Si) surface contamination after post-CMP SC-1 cleaning and no modification of the BPZT cation surface composition. The formation of BaOH at the surface was observed after cleaning, which had however no effect on the BPZT properties. After CMP and cleaning, BPZT showed minor degradations of the ferroelectric and piezoelectric responses (on the order of 10-20%), especially for thinner films. The minor degradation could be attributed to the amorphization of the top BPZT surface and the formation of a thin dead layer. Moreover, CMP and cleaning had only minor effects on the microwave dielectric properties up to 30 GHz, with a small reduction of the effective permittivity due to the dead layer but no impact on the "bulk" permittivity and the frequency dispersion (dielectric relaxation).

In terms of final roughness and removal rate, as well as ferroelectric, piezoelectric, and dielectric characteristics, the approach provides thus excellent results, also compared to previous reports[24,22]. The polishing and cleaning procedures show limited impact on the BPZT properties at microwave frequencies, which renders them suitable in applications of BPZT, *e.g.*, in ultrasound transducers or multiferroic devices. While a small degradation due to the surface dead layer has been observed, its impact can be expected to be rather limited, especially given the potential significant improvement obtained from the reduction of the surface roughness.

**Acknowledgments**

This work has been supported by imec's industrial affiliate program on beyond CMOS logic. It has been partially funded by the European Union's Horizon 2020 research and innovation program within the European Innovation Council Pathfinder FET-OPEN project CHIRON under grant agreement No. 801055. F.L. acknowledges support by the Research Foundation Flanders (FWO) through a PhD fellowship under grant agreement No. 1183722N.

**References**

1. N. Korobova, in *Advanced Piezoelectric Materials*, p. 533–574, Elsevier (2017) https://linkinghub.elsevier.com/retrieve/pii/B978008102135400014X.

2. H. Bhugra and G. Piazza, Eds., *Piezoelectric MEMS Resonators*, Springer International Publishing, Cham, (2017) http://link.springer.com/10.1007/978-3-319-28688-4.

3. B. Jaffe, R. S. Roth, and S. Marzullo, *Journal of Applied Physics*, **25**, 809–810 (1954).

4. M. Ahart, M. Somayazulu, R. E. Cohen, P. Ganesh, P. Dera, H. K. Mao, R. J. Hemley, Y. Ren, P. Liermann, and Z. Wu, *Nature*, **451**, 545–548 (2008).

5. S. Uhlig, M. Nicolai, A. Schönecker, and A. Michaelis, *Materials Science and Technology*, **25**, 1321–1324 (2009).

6. G. L. Smith, J. S. Pulskamp, L. M. Sanchez, D. M. Potrepka, R. M. Proie, T. G. Ivanov, R. Q. Rudy, W. D. Nothwang, S. S. Bedair, C. D. Meyer, and R. G. Polcawich, D. J. Green, Editor. *J. Am. Ceram. Soc.*, **95**, 1777–1792 (2012).

7. W. Kang, J. Jung, W. Lee, J. Ryu, and H. Choi, *J. Micromech. Microeng.*, **28**, 075015 (2018).

8. Y. Chen, X. Bao, C. M. Wong, J. Cheng, H. Wu, H. Song, X. Ji, and S. Wu, *Ceramics International*, **44**, 22725–22730 (2018).

9. Z. Jiang, R. J. Dickinson, T. L. Hall, and J. J. Choi, *IEEE Trans. Ultrason., Ferroelect., Freq. Contr.*, **68**, 2164–2171 (2021).

10. N. Yamauchi, T. Shirai, T. Yoshihara, Y. Hayasaki, T. Ueda, T. Matsushima, K. Wasa, I. Kanno, and H. Kotera, *Appl. Phys. Lett.*, **94**, 172903 (2009).

11. U. K. Bhaskar, D. Tierno, G. Talmelli, F. Ciubotaru, C. Adelmann, and T. Devolder, *IEEE Trans. Ultrason., Ferroelect., Freq. Contr.*, **67**, 1284–1290 (2020).

12. G. Srinivasan, *Annu. Rev. Mater. Res.*, **40**, 153–178 (2010).

13. C. A. F. Vaz, J. Hoffman, C. H. Ahn, and R. Ramesh, *Advanced Materials*, **22**, 2900–2918 (2010).

14. R. Duflou, F. Ciubotaru, A. Vaysset, M. Heyns, B. Sorée, I. P. Radu, and C. Adelmann, *Appl. Phys. Lett.*, **111**, 192411 (2017).

15. R.-C. Peng, J.-M. Hu, L.-Q. Chen, and C.-W. Nan, *NPG Asia Mater*, **9**, e404–e404 (2017).

16. Z. Ren, M. Wang, P. Liu, Q. Liu, K. Wang, G. Jakob, J. Chen, K. Meng, X. Xu, J. Miao, and Y. Jiang, *Adv. Electron. Mater.*, **6**, 2000102 (2020).

17. F. Wang, M. Zhu, Y. Qiu, R. Guo, G. Wu, G. Yu, and H. Zhou, *IEEE Trans. Electron Devices*, **69**, 1650–1657 (2022).

18. A. Mahmoud, F. Ciubotaru, F. Vanderveken, A. V. Chumak, S. Hamdioui, C. Adelmann, and S. Cotofana, *Journal of Applied Physics*, **128**, 161101 (2020).

19. V. Porokhonskyy and D. Damjanovic, *Appl. Phys. Lett.*, **94**, 212906 (2009).

20. D. Min, N. Hoivik, and U. Hanke, *J Electroceram*, **28**, 53–61 (2012).

21. R. Zachariasz, D. Bochenek, K. Dziadosz, J. Dudek, and J. Ilczuk, *Archives of Metallurgy and Materials*, **56** (2011) http://journals.pan.pl/dlibra/publication/99781/edition/86079/content.

22. D. Tierno, M. Dekkers, P. Wittendorp, X. Sun, S. C. Bayer, S. T. King, S. Van Elshocht, M. Heyns, I. P. Radu, and C. Adelmann, *IEEE Trans. Ultrason., Ferroelect., Freq. Contr.*, **65**, 881–888 (2018).

23. M. D. Nguyen, R. Tiggelaar, T. Aukes, G. Rijnders, and G. Roelof, *J. Phys.: Conf. Ser.*, **922**, 012022 (2017).

24. Y.-J. Seo, J.-S. Park, and Woo-Sun Lee, *Microelectronic Engineering*, **83**, 2238–2242 (2006).

25. N.-H. Kim, Y.-K. Jun, P.-J. Ko, and W.-S. Lee, *Journal of Vacuum Science & Technology A: Vacuum, Surfaces, and Films*, **26**, 720–723 (2008).

26. N.-H. Kim, P.-J. Ko, and W.-S. Lee, in *2008 17th IEEE International Symposium on the Applications of Ferroelectrics,*, p. 1–2, IEEE, Santa Re, NM, USA (2008) http://ieeexplore.ieee.org/document/4693729/.

27. S. H. Choi, B. J. Bae, Y. H. Son, J. E. Lim, D. C. Yoo, D. H. Im, J. E. Heo, S. D. Nam, J. H. Park, C. K. Hong, H. K. Cho, and J. T. Moon, *Integrated Ferroelectrics*, **75**, 215–223 (2005).

28. S. H. Choi, H. Y. Ko, J. E. Heo, Y. H. Son, B. J. Bae, D. C. Yoo, D. H. Im, Y. J. Jung, K. R Byun, J. H. Hahm, S. H. Shin, B. U. Yoon, C. K. Hong, H. K. Cho, and J. T. Moon, *Integrated Ferroelectrics*, **84**, 147–158 (2006).

29. S. Wang, B. Ma, J. Deng, H. Qu, and J. Luo, *Microsystem Technologies*, **21**, 1053–1059 (2015).

30. Q. Zou, H. E. Ruda, and B. G. Yacobi, *Appl. Phys. Lett.*, **78**, 1282–1284 (2001).

31. T. Kobayashi, M. Ichiki, R. Kondou, K. Nakamura, and R. Maeda, *J. Micromech. Microeng.*, **18**, 035007 (2008).

32. Z. Ma, A. J. Becker, P. Polakos, H. Huggins, J. Pastalan, H. Wu, K. Watts, Y. H. Wong, and P. Mankiewich, *IEEE Trans. Electron Devices*, **45**, 1811–1816 (1998).

33. A. Thøgersen, J. H. Selj, and E. S. Marstein, *J. Electrochem. Soc.*, **159**, D276–D281 (2012).

34. J. Hossain, K. Kasahara, D. Harada, A. T. M. Saiful Islam, R. Ishikawa, K. Ueno, T. Hanajiri, Y. Nakajima, Y. Fujii, M. Tokuda, and H. Shirai, *J. Appl. Phys.*, **122**, 055101 (2017).

35. K. Ramam and M. Lopez, *J. Phys. D: Appl. Phys.*, **39**, 4466–4471 (2006).

36. B. K. Bammannavar and L. R. Naik, *Smart Mater. Struct.*, **18**, 085013 (2009).

**Tables**

| Condition | Table/Head speed | Downforce | RR | RMS |
|---|---|---|---|---|
| 1 | Low | Low | (1.61±0.05) nm/s | 0.79±0.15 nm |
| 2 | Low | High | (2.16±0.02) nm/s | 1,04±0.15 nm |
| 3 | High | Low | (2.72±0.03) nm/s | 0.86±0.08 nm |
| 4 | High | High | (3.19±0.02) nm/s | 1.19±0.09 nm |

Table 1 – Parameters used for the different conditions of CMP process under investigations and results for removal rate (RR) and RMS roughness for BPZT.

| | Ba3d5 | Pb4f | Zr3d | Ti2p |
|---|---|---|---|---|
| As deposited | 3.5±0.3 | 53±5 | 26±3 | 17±2 |
| After CMP | 3.0±0.3 | 51±5 | 26±3 | 20±2 |
| After CMP + SC-1 | 3.7±0.3 | 53±5 | 29±3 | 15±2 |

Table 2 - Atomic concentration (at.%) for BPZT components (Ba, Pb, Zr and Ti) under different process conditions; error given by the accuracy of the inspection (around 10%)

| | Thickness (nm) | $P_{sat+}$ ($\mu C/cm^2$) | $P_{sat-}$ ($\mu C/cm^2$) | $E_{c+}$ (KV/cm) | $E_{c-}$ (KV/cm) | $d_{33}$ (pm/V) |
|---|---|---|---|---|---|---|
| After CMP | 200 | 37.1 | -37.3 | 50 | -118 | 48 |
| As depos. | | 42.1 | -42.0 | 42.0 | -106 | 65 |
| After CMP | 450 | 39.1 | -39.1 | 24.7 | -62.1 | 96 |
| As depos. | | 42.7 | -42.6 | 24.9 | -51.1 | 115 |

Table 3 - Ferroelectric and piezoelectric properties of 200nm and 450nm thick BPZT thin films, comparison for BPZT after CMP and as deposited

| | Slope $1/\varepsilon_{BPZT}$ | $\varepsilon_{BPZT}$ | Intercept $t_{DL}/\varepsilon_{DL}$ |
|---|---|---|---|
| BPZT as deposited | 6.94±0.04 E-4 | 1441±8 | (0.070±0.002) nm |
| BPZT after CMP | 7.2±0.5 E-4 | 1390±90 | (0.078±0.027) nm |

Table 4 - Intrinsic permittivity of BPZT (with and without CMP) and normalized effective permittivity of the dead layer extracted by the linear fit

**Figure Captions**

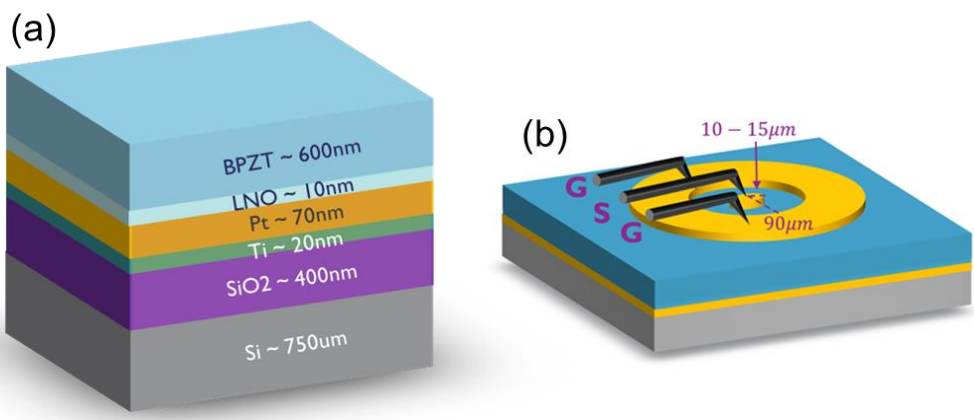

*Figure 1 – Schematic representation of (a) the stack deposited and (b) of concentric capacitors design.*

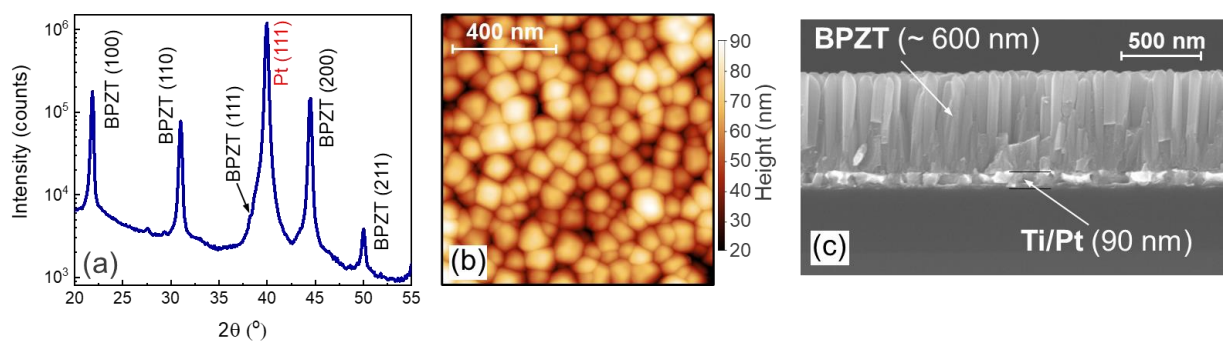

*Figure 2 - (a) 2θ- ω XRD pattern, (b) atomic force microscopy surface analysis and (c) cross-sectional scanning electron micrographs of 600nm thick BPZT films*

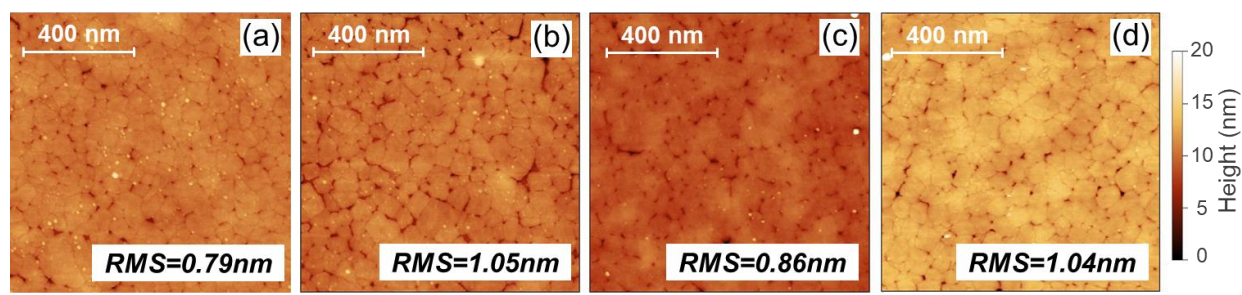

*Figure 3 – Atomic force microscopy surface analysis of BPZT after different conditions of CMP. Comparison of (a) the best condition (condition 1, low speeds and low down force) for 2 minutes with (b) longer polishing (3 minutes); (c) high speeds (condition 3) and (d) high down force (condition 2).*

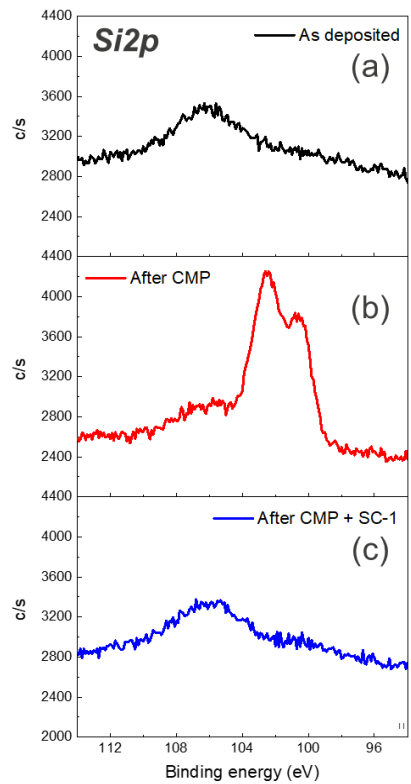

*Figure 4 - XPS scans of Si2p on the surface of the BPZT thin films as a function of the process: BPZT as deposited (a), after CMP (b) and after CMP + cleaning in SC-1 (c)*

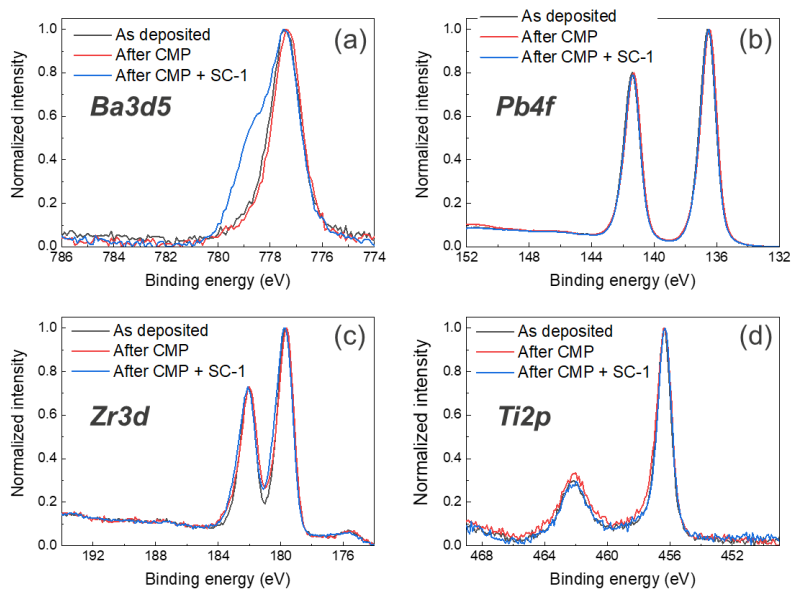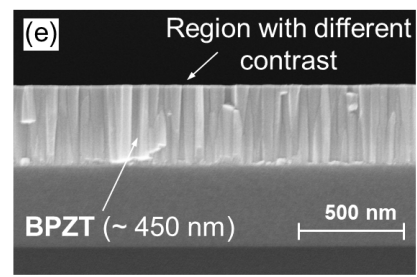

*Figure 5 - XPS scans of (a) Ba3d5, (b) Pb4f, (c) Zr3d, and (d) Ti2p of the BPZT thin films as a function of the process: BPZT as deposited (black curve), after CMP (red curve) and after CMP + cleaning in SC-1 (blue curve). (e) X-SEM image of 450nm thick BPZT after CMP process.*

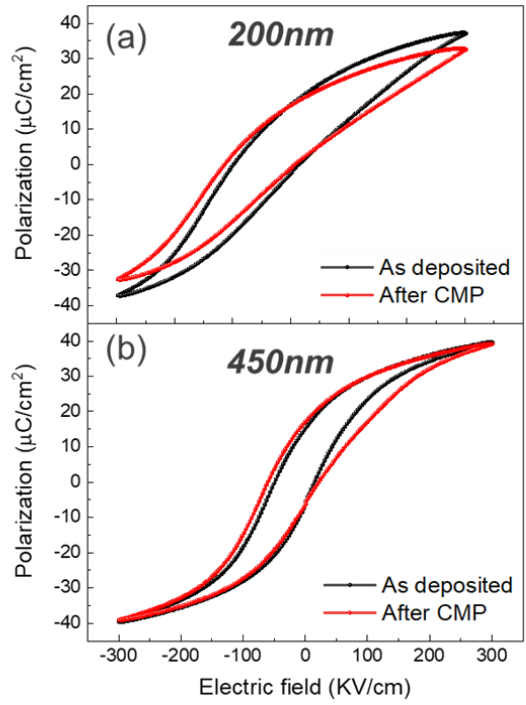

*Figure 6 - Comparison P-E ferroelectric loops of BPZT as deposited (in black) and after CMP (in red) for (a) 200nm thick and (b) 450nm thick films.*

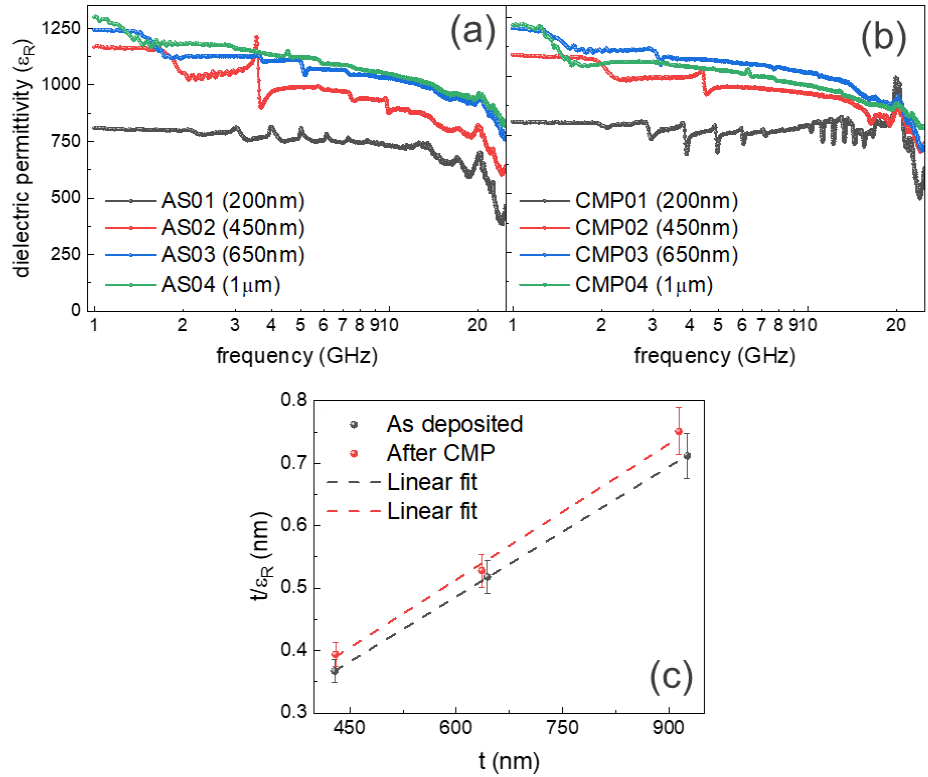

*Figure 7 - Dielectric permittivity $\varepsilon_r$ of BPZT films (a) as deposited and (b) after CMP with thicknesses as indicated. (c) Linear fit of the reciprocal normalized effective permittivity vs BPZT thickness at 1GHz, with (in red) and without (in black) CMP process.*